\documentstyle[prb,aps,epsfig]{revtex}
\begin{document}
\title{Concentration-Pressure phase diagram for rich Zr PZT ceramics}
\author{A. G. Souza Filho\thanks{
Corresponding Author: Tel: +55 85 2889912, Fax: +55 85 2874138 (A.
G. Souza Filho)}, P. T. C.
Freire\thanks{e-mail:agsf@fisica.ufc.br}, A. P. Ayala, J. M.
Sasaki, I. Guedes, J. Mendes Filho, F. E. A. Melo}
\address{Departamento de F\'{\i}sica, Universidade Federal do
Cear\'{a}, Caixa Postal 6030,\\ 60455-760 Fortaleza, Cear\'{a},
Brazil}
\author{E. B. Ara\'ujo and J. A. Eiras}
\address{Departamento de F\'{\i}sica, Universidade Federal de S\~ao Carlos,
Caixa Postal 676,\\ 13565-670 S\~ao Carlos, S\~ao Paulo, Brazil}
\date{\today}
\maketitle
\begin{abstract}
This work reports on the systematic high pressure Raman studies
in the PbZr$_{1-x}$Ti$_x$O$_3$ ($0.02 \leq x \leq 0.14$) ceramics
performed at room temperature. The pressure dependence of the
Raman spectra reveals the stable phases of the material under
pressure variation. The results allowed us to propose a
concentration-pressure phase diagram for rich Zr PZT system up to
pressures of 5.0 GPa.
\end{abstract}

\bigskip



Lead Zirconate titanate, PbZr$_{1-x}$Ti$_x$O$_3$ is widely known
for their technological importance in the field of electronic,
sensors, and non-volatile ferroelectric memory devices. Due this,
PZT system at different forms (ceramics, single crystals and thin
films) is one of the most studied ferroelectric material for over
50 years by several experimental techniques such as X-ray and
neutron diffraction\cite{ap,hz,hf,dlc}, electric
measurements\cite{zu,jh}, and Raman
spectroscopy\cite{ieh,db,dca,kr,gb,bas,mj,rsk}. Their dielectric,
piroelectric, and ferroelectric properties are strongly dependent
both on the structural phase and on the preparation
method\cite{rsk}. Depending on the $x$ value PZT exhibits at
atmospheric pressure and room temperature different phases as
follows\cite{bj}. For 0 $\leq x \leq$ 0.05, PZT presents an
orthorhombic antiferroelectric structure belonging to the
C$_{2v}^{8}$ space group. For $x$ varying from 0.05 to 0.37 (0.37
to 0.48) PZT presents rhombohedral ferroelectric low temperature
phase F$_R$(LT) (rhombohedral ferroelectric high temperature phase
F$_R$(HT)) belonging to the space group C$_{3v}^{5}$
(C$_{3v}^{6}$). The composition around x = 0.48 defines a region
known as morphotropic phase boundary (MPB) which divides the
rhombohedral from the tetragonal phases. From x = 0.48 to x = 1.0
the PZT exhibits a tetragonal structure belonging to the space
group C$_{4v}^{1}$. Recently, new features on the MPB region were
reported\cite{agsf1,bn,jag}. Noheda et al.\cite{bn,jag} using
high-resolution synchroton X-ray powder diffraction and dielectric
measurements, a new monoclinic ferroelectric phase belonging to
the C$_s^2$ space group was discovered at low temperatures.

There are several number of theoretical and experimental efforts
in order to determine the thermodynamically stable phase of PZT
when pressure varies. Cerdeira et al\cite{cerd} have studied the
behavior of PbTiO$_3$ single crystal up to 80 kbar. The pressure
dependence of the lowest E(TO) soft mode frequency obeys the
Curie-Weiss law which predict that PbTiO$_3$ undergoes a
structural phase transition from a ferroelectric tetragonal to a
cubic paraelectric phase at pressure of about 90 kbar. Bauerle et.
al \cite{wbh} have studied PbTi$_{0.10}$Zr$_{0.90}$O$_3$ ceramics
by Raman spectroscopy with pressures up to 6.85 GPa. They showed
that the material undergoes a phase transition at 0.57 GPa from
the initial room temperature-atmospheric pressure (F$_R$(LT)) to a
high-temperature rhombohedral phase. Between 0.8 and 0.91 GPa,
PbTi$_{0.10}$Zr$_{0.90}$O$_3$ goes to the orthorhombic
antiferroelectric phase and between 3.97 and 4.2 GPa a new phase
is reached, with a symmetry higher than that of the
antiferroelectric phase.

Recently, by means of dielectric, X-ray and Raman measurements,
Furuta {\it et al.}\cite{yb,se} showed that a PbZrO$_{3}$
polycrystalline fine-powder sample undergoes a rich phase
transition sequence up to 30 GPa: from the antiferroelectric phase
to an orthorhombic phase I' at 2.3 GPa, from an orthorhombic phase
I' to an orthorhombic phase I'' at 17.5 GPa, and finally from an
orthorhombic phase I'' to a monoclinic phase at 23 GPa. More
recently, Souza Filho et al.\cite{agsf} have studied the
PbZr$_{0.94}$Ti$_{0.06}$O$_3$, which besides presenting at room
temperature the same phase that of
PbZr$_{0.90}$Ti$_{0.10}$O$_3$\cite{wbh}, it presents a sequence of
phase transitions very different from the latter. This fact points
out to the richness of the PZT concentration-pressure phase
diagram.

In spite of the variety of studies of PZT under temperature
variation by several techniques, there is a limited number of
Raman studies performed on the PZT system under pressure
variation. The micro-Raman spectroscopy, which is quite useful to
investigate a localized area in the probed sample with a spatial
resolution of the order of $\mu$m, is one of the most powerful
technique to investigate phase transition in condensed matter
under pressure variation. The purpose of this work is to
investigate through micro-Raman spectroscopy the structural
properties of PbZr$_{1-x}$Ti$_x$O$_3$ ceramics under high
hydrostatic pressure. A careful analysis of the Raman spectra of
samples with six different $x$ value yielded information
concerning the different stable phases of PbZr$_{1-x}$Ti$_x$O$_3$
under pressure variation. Based on previous Raman and X-ray
investigations in PbZr$_{1-x}$Ti$_x$O$_3$ and PbZrO$_3$ we propose
a concentration-pressure phase diagram for rich Zr PZT system up
to pressures of 5.0 GPa for $0.02 \leq x \leq 0.14$.

The preparation of our samples is described elsewhere\cite{agsf}.
Raman microprobe spectroscopy experiments were performed at room
temperature in the backscattering geometry using the
514.5\thinspace nm radiation line of a Ar-ion laser for
excitation. The backscattered light was analyzed using a Jobin
Yvon Triplemate 64000. A $N_2$- cooled Charge Coupled Device (CCD)
detector was used to detect the Raman signal. The spectrometer
slits were set for a 2\thinspace cm$^{-1}$ spectral resolution. An
Olympus microscope lens and an objective with a numerical aperture
NA $=0.80$ were employed to focus the laser beam at the polished
sample surface. The laser power impinging on the samples surface
was of the order of 10\thinspace mW. The pressure transmitting
fluid used was 4:1 methanol-ethanol and pressure calibration was
achieved with the well known pressure shift of the ruby
luminescence lines.

\noindent{\bf Orthorhombic Antiferroelectric Phase~-~\bf
PbZr$_{1-x}$Ti$_x$O$_3$ with $\bf 0.02 \leq x \leq 0.04$}~-~
Figure 1 shows the unpolarized Raman spectra for
PbZr$_{0.98}$Ti$_{0.02}$O$_3$. At room temperature and room
pressure, this composition has an orthorhombic structure belonging
to the space group C$_{2v}^{8}$ as PbZrO$_3$\cite{pasto}. Below
150 cm$^{-1}$ (spectral region depicted in the insert of Fig. 1)
six modes at frequencies of 35, 44, 50, 55, 70, and 132 $cm^{-1}$
are observed. This spectral region contains the external modes
related to Pb- lattice modes\cite{car92}. In the high frequency
region, $150 \leq \omega \leq 1000$ cm$^{-1}$ some internal modes
related to certain polyatomic groups of the material appears in
the Raman spectra. In the 0.0 GPa spectrum, bands at 204, 232
(Zr-O bending), 285, 330, and 344 ($ZrO_{3}$ torsions), 501, and
532 cm$^{-1}$ (Zr-O stretching) are also observed. The assignments
of these bands were made based on the works in PbZrO$_{3}$ single
crystals\cite{pasto}.

To understand the pressure dependence of Raman spectra for
PbZr$_{0.98}$Ti$_{0.02}$O$_3$, let us remember the main results
obtained by Furuta et. al.\cite{se} in PbZrO$_3$ polycrystalline
up to pressure of 5.0 GPa. These authors showed that the left-hand
side mode of doublet located at about 210 cm$^{-1}$ disappears at
pressures higher then 2.3 GPa and that the right side mode
increases in intensity. This spectral discontinuity is attributed
to the orthorhombic(I) antiferroelectric phase $\rightarrow$
orthorhombic(I') phase transition already determined by means of
high-pressure X-ray diffraction measurements. For
PbZr$_{0.98}$Ti$_{0.02}$O$_3$ the doublet mode is characterized by
bands at 207 (labeled with {\bf a} in the Fig. 1) and 232
cm$^{-1}$ (labeled with {\bf b} in the Fig. 1). Upon increasing
pressures, the spectral features for the doublet is the same that
was found in PbZrO$_3$. This spectral discontinuity observed for
PbZr$_{0.98}$Ti$_{0.02}$O$_3$ around 2.18 GPa indicates the
orthorhombic(I) antiferroelectric phase $\rightarrow$
orthorhombic(I') phase transition. Moreover, the transition can be
clearly identified by drastic changes in the lattice mode region
(insert in Fig. 1), in particular through the observation of a
band (marked with an arrow) that disappears in the spectra of high
pressure phase.

For pressures above 2.18 GPa and up to 4.0 GPa , the Raman spectra
remain the same which could suggest that the material did not
undergoes additional structural phase transitions as occur for the
PbZrO$_{3}$\cite{yb,se}. The Raman spectra of
PbZr$_{0.96}Ti_{0.04}O_{3}$ are qualitatively similar to those of
PbZr$_{0.98}Ti_{0.02}O_{3}$. The only difference is that the
pressure where the phase transition orthorhombic phase (I)
$\rightarrow$ orthorhombic phase (I') occurs is 2.4 for the
PbZr$_{0.96}Ti_{0.04}O_{3}$.

\noindent{\bf Rhombohedral Ferroelectric Phase ~~-~~\bf
PbZr$_{1-x}$Ti$_x$O$_3$ with $\bf 0.10 \leq x \leq 0.14$}~-~ This
set of samples presents at room temperature and atmospheric
pressure a rhombohedral structure belonging to the C$_{3v}^6$
space group\cite{bj}. First, let us describe the pressure
dependence of Raman spectra for PbZr$_{0.94}Ti_{0.06}O_{3}$ and
PbZr$_{0.92}Ti_{0.08}O_{3}$. The former was the subject of our
recent work\cite{agsf} where we have shown that
$PbZr_{0.94}Ti_{0.06}O_{3}$ undergoes two different phase
transitions up to 3.7 $GPa$ as follow:
rhombohedral(LT)$\stackrel{0.3GPa}{\rightarrow }$orthorhombic(I)
$\stackrel{2.9GPa}{\rightarrow}$orthorhombic(I$^{\prime}$). These
phase transitions and the presssure dependence of Raman active
modes were described in details elsewhere\cite{agsf}. It should be
pointed that composition PbZr$_{0.92}Ti_{0.08}O_{3}$ have exactly
the same pressure dependence and the transitions rhombohedral(LT)
$\rightarrow$ orthorhombic(I) $\rightarrow$
orthorhombic(I$^{\prime }$) occur at 0.5  and 3.4 GPa,
respectively.

Figure 2 shows the Raman spectra for PbZr$_{0.90}$Ti$_{0.10}$O$_3$
recorded at different pressures. Ba$\ddot u$erle et. al.
\cite{wbh} have reported for PbZr$_{0.90}$Ti$_{0.10}$O$_3$
composition the following sequence of pressure-induced phase
transitions:
\noindent{rhombohedral(LT)$\stackrel{0.57GPa}{\rightarrow}$
rhombohedral (HT) $\stackrel{0.8GPa}{\rightarrow }$
orthorhombic(I) $\stackrel{3.9GPa}{\rightarrow}$higher symmetry
phase (probably cubic). Our results agree in part with those
reported in Ref. 20: in the pressure range of 0.0 - 1.0 GPa, we
found the same sequence of phase transitions. Unfortunately,
Ba$\ddot u$rele et al. \cite{wbh} did not report the spectral
region above 200 $cm^{-1}$. By observing the double bands (labeled
with a and b) of the spectra in the insert of Figure 2, it is
clear that the stable phase above 4.2 GPa for
PbZr$_{0.90}$Ti$_{0.10}$O$_3$ is the orthorhombic phase I' as
already depicted in this paper and in the works of Refs. 22 and
23. Thus, the transition observed in PbZr$_{0.90}$Ti$_{0.10}$O$_3$
\cite{wbh} at 3.9 GPa is from an antiferroelectric orthorhombic
phase to the orthorhombic I' instead of antiferroelectric
orthorhombic to a paraelectric cubic phase as proposed by Ba$\ddot
u$erle et al. \cite{wbh}.

For the compositions PbZr$_{0.86}$Ti$_{0.14}$O$_3$ (see Fig. 3)
and PbZr$_{0.88}$Ti$_{0.12}$O$_3$ it was observed the almost the
same qualitative features that was found for
PbZr$_{0.90}$Ti$_{0.10}$O$_3$. It was found an increase in the
rhombohedral(LT) $\rightarrow$ rhombohedral(HT) $\rightarrow$
orthorhombic(I) phase transition pressures and surprisingly a
decreasing in the orthorhombic(I) $\rightarrow$ orthorhombic(I')
transition pressure. Finally, the results of all compositions
investigated in this work can be summarized in the equilibrium
phase diagram (concentration-pressure) for Zr rich PZT ceramics
depicted in Fig. 4.

In conclusion, we have performed a systematic high-pressure Raman
study on several PbZr$_{1-x}$Ti$_x$O$_3$ samples in order to
reveal the effects of the pressure on the stable phases of Zr rich
PZT ceramics. We detected the existence of a triple point which
limit the antiferroelectric, rhombohedral low temperature phase,
and rhombohedral high temperature phase for Ti concentration
around 0.09. Also, it should be pointed out the possibility of the
existence of another triple point in the phase diagram that would
be determined by the extension limit of antiferroelectric phase.

\bigskip

\noindent{\bf Acknowledgments}~~-~~A.G.S.F. acknowledges the
fellowship received from Funda\c{c}\~ao Cearense de Amparo \`a
Pesquisa (FUNCAP). P.T.C.F. acknowledges FUNCAP for grant
n$^{\b{o}}$ 017/96 PD. Financial support from CNPq, FAPESP and
FINEP, Brazilian funding agencies, is also grateful acknowledged.

\begin{figure}
\centerline{\epsfig{file=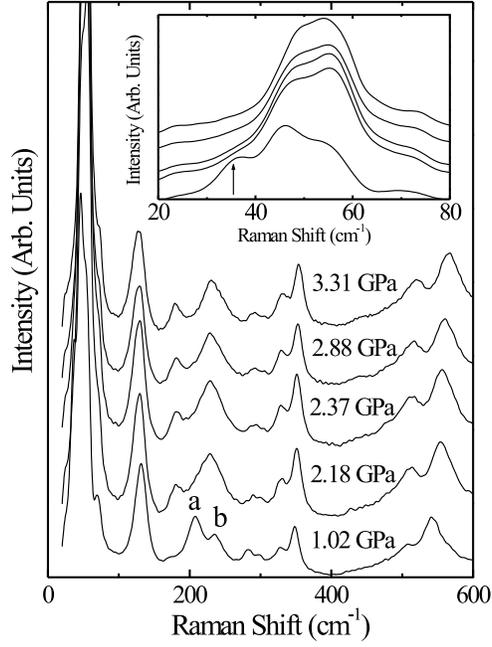,width=18pc}} \caption{Raman
spectra for PbZr$_{0.98}$Ti$_{0.02}$O$_3$ ceramics recorded at
pressures up to 5.0 GPa. The insert depicted the external region
mode.}
\end{figure}

\begin{figure}
\centerline{\epsfig{file=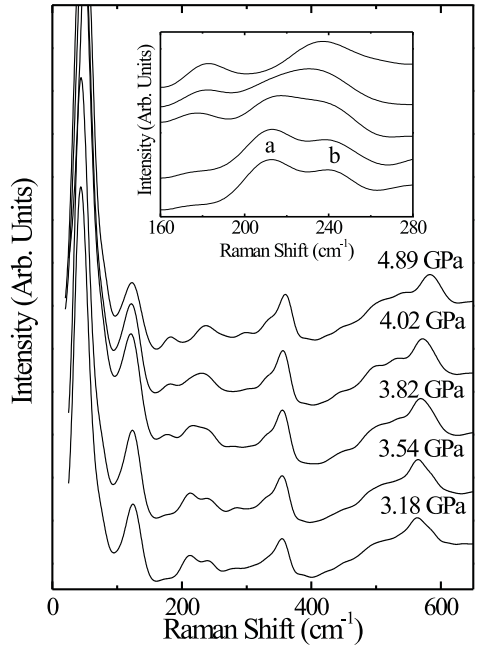,width=18pc}} \caption{Raman
spectra for PbZr$_{0.90}$Ti$_{0.10}$O$_3$ ceramics recorded at
pressures up to 5.0 GPa. The insert show that a doublet mode
behavior indicating a orhthorhombic(I) $\rightarrow$
orhthorhombic(I') phase transition.}
\end{figure}

\begin{figure}
\centerline{\epsfig{file=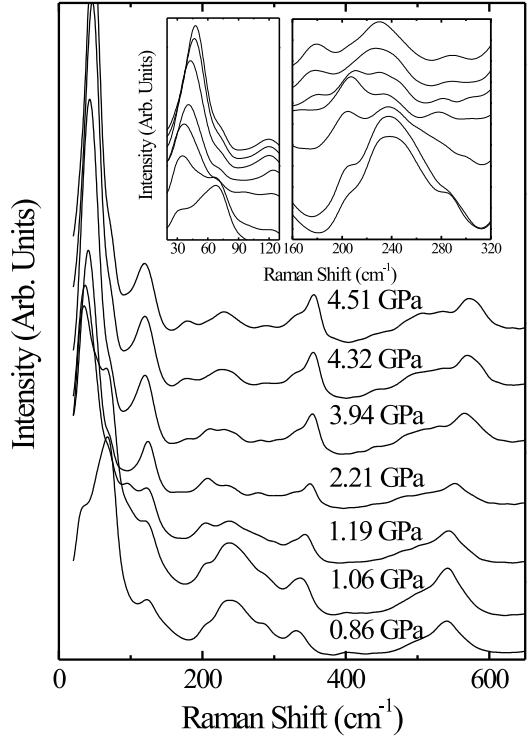,width=20pc}} \caption{Raman
spectra for PbZr$_{0.86}$Ti$_{0.14}$O$_3$ ceramics recorded at
pressures up to 5.0 GPa.}
\end{figure}

\begin{figure}
\centerline{\epsfig{file=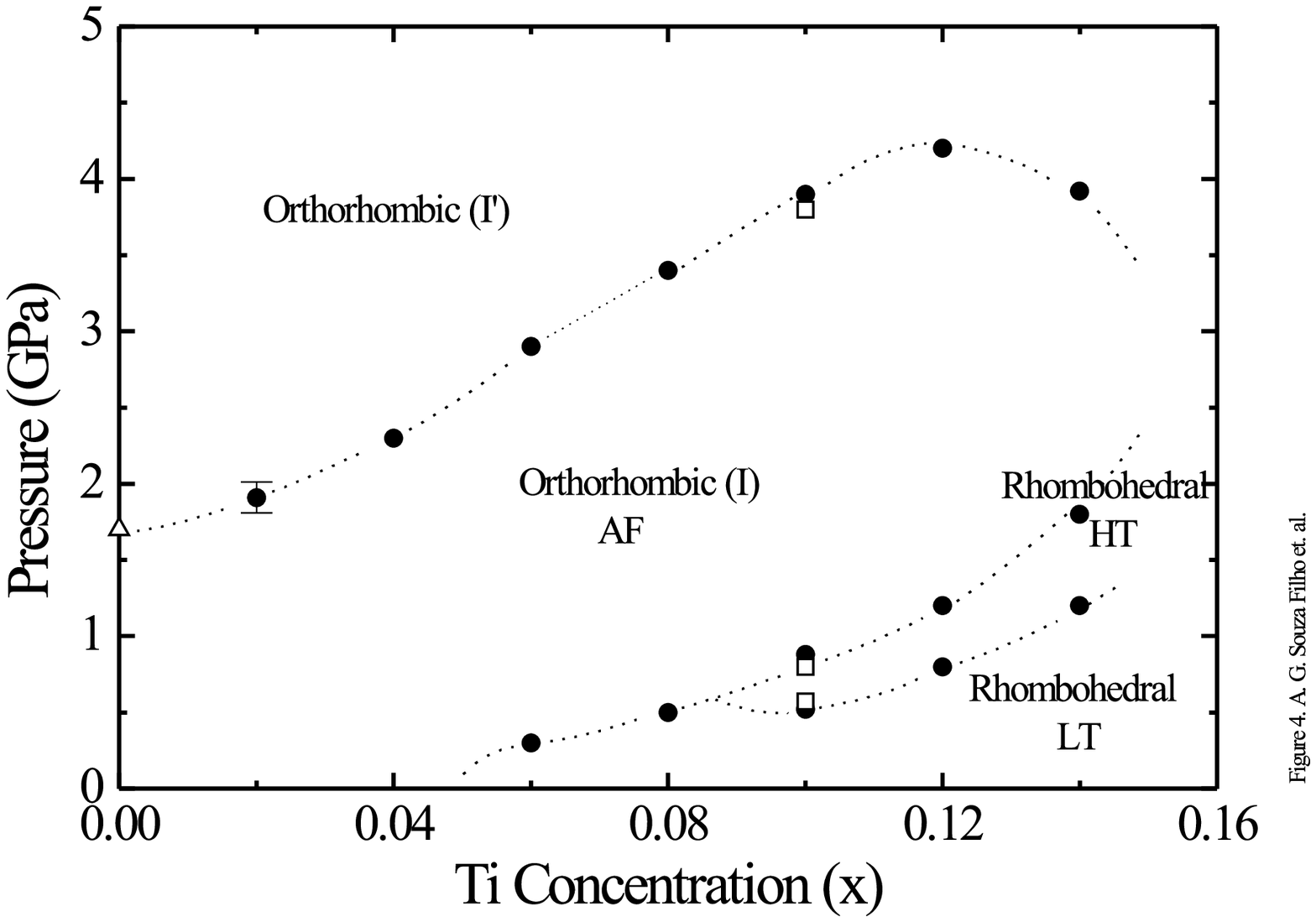,width=30pc}} \caption{Plot of
Concentration-Pressure phase diagram for Zr rich PZT system. The
open triangle and open squares are data taken from Furuta et
al.\cite{se} and Ba$\ddot u$erle et al.\cite{wbh}, respectively.
For all points, the error bar was of order of 0.15 GPa. The dashed
lines are visual guides.}
\end{figure}

\end{document}